\documentclass[aps,pra,reprint,superscriptaddress]{revtex4-2}
\usepackage{graphicx}
\usepackage{amsmath}
\usepackage{xcolor}
\usepackage{braket}
 \usepackage{amssymb}
\usepackage[pdftex]{hyperref} 
\begin{document}
\preprint{}
\title{Ion-laser-like interaction in optomechanical systems with Kerr nonlinearities}
\author{I. Ramos-Prieto}
\email[e-mail:]{ iranrp123@gmail.com}
 \affiliation{Instituto de Ciencias Físicas, Universidad Nacional Autónoma de México, Apdo. Postal 48-3, Cuernavaca, Morelos 62251, Mexico}
\author{R. Román-Ancheyta}
\affiliation{Instituto Nacional de Astrofísica Óptica y Electrónica, Calle Luis Enrique Erro No. 1, Santa María Tonantzintla, Pue., 72840, Mexico}
\author{J. Récamier}
 \affiliation{Instituto de Ciencias Físicas, Universidad Nacional Autónoma de México, Apdo. Postal 48-3, Cuernavaca, Morelos 62251, Mexico}
\author{M. Berrondo}
 \affiliation{Department of Physics and Astronomy, Brigham Young University, Provo, UT 84602, USA}
\author{H. M. Moya-Cessa}
\affiliation{Instituto Nacional de Astrofísica Óptica y Electrónica, Calle Luis Enrique Erro No. 1, Santa María Tonantzintla, Pue., 72840, Mexico}
\date{\today}
\begin{abstract}
Introducing a Kerr medium in a cavity coupled to a harmonically moving mirror, we reproduce known and solvable interactions such as two-coupled harmonic oscillators and ion-laser like interactions for specific conditions. This is achieved by a unitary transformation allowing us to tune off the Kerr medium in order to simplify the Hamiltonian.
\end{abstract}
\maketitle
\section{Introduction}
Cavity optomechanics deals with the coupling of an optical cavity field with a mechanical oscillator, symbiotically generating a joint response of the displacement of the mechanical oscillator and the resonant frequency of the cavity \cite{Aspelmeyer_2014}. The ability to observe and manipulate quantum  behavior of macroscopic objects has generated a wide range of theoretical and experimental approaches, among them: nonclassical states \cite{Mancini_1997,Bose_1997,Marshall_2003,Xu_2013,Li_2018}, weak and strong coupling regime and blocked photons  \cite{Groblacher_2009,Nunnenkamp_2011,Mikkelsen_2017,Kumar_2010,Rabl_2011}, optical bistability \cite{Dorsel_1983,Gozzini_1985,Shahidini_2014,Ghobadi_2011,Aldana_2013}, and control of the dynamics of the micro-mirror enhancing the Kerr effect \cite{Gong_2009,Huang_2009,Huang_2020}.

An important challenge in this field is the way to prepare or bring the mechanical oscillator to its ground state \cite{Braginsky_2002}.  With the advent of resolved sideband cooling \cite{Neuhauser_1978,Wineland_1979}, radiation pressure-driven mechanical oscillators have followed a similar path \cite{Marquardt_2007,Kippenberg_2008,Schliesser_2008,Teufel_2008,Teufel_2011,Chan_2011,Clark_2017}. Indeed, an analogy has been found between the sideband cooling mechanism of the trapped ions in the Lamb-Dicke regime and cooling via dynamical backaction \cite{Wilson_2007}. Furthermore, the Hamiltonian of a pumped optomechanical system has been shown to be quasi-analogous to the Hamiltonian of a trapped ion oscillating in one dimension \cite{Ventura_2015}, since an effective Kerr term prevails, {\it i.e.,} the cavity  mode plays the role of pseudo-spin operators. An additional characteristic of an optomechanical system is the fact that the coupling turns out to be inherently nonlinear because the frequency (through the cavity length) is modified by the radiation pressure \cite{Law_1995}. It is relevant to note that there is an evolution analogous to that of a Kerr-like nonlinearity \cite{Mancini_1997,Bose_1997}. In particular there is an equivalence between an optomechanical system and a Kerr medium \cite{Aldana_2013}, and also, the Kerr medium enhances and simplify the thermometry procedure in the nonlinear optomechanical regime~\cite{Montenegro_2020}, and improve quantum synchronization in coupled optomechanical systems \cite{Qiao_2018}. However, the effective Kerr medium precludes simpler and more approximate handling over the pumped optomechanical system. This difficulty is an important motivations of this Letter.

In this Letter, we show that the optomechanical interaction in a Kerr medium tuned to a specific value that cancels the effective Kerr term produced by a transformation of the mirror-field Hamiltonian, leads to two different kinds of equivalent physical systems described below. We have not included dissipation in any of the subsystems in order to concentrate on the central idea that we want to highlight.
\section{Model}
We consider a single-mode quantized electromagnetic field in a cavity including a Kerr medium characterized by the constant $\chi$ \cite{Buzek_1990}, and allow one of its mirrors to oscillate harmonically. This system may be described by the optomechanical Hamiltonian in a frame rotating at the laser frequency $\omega_d$ (with $\hbar=1$),
\begin{equation}\label{mH}
	\begin{split}
		\hat{\mathcal{H}}&=\hat{\mathcal{H}}_0+\hat{\mathcal{H}}_d,\\
		\hat{\mathcal{H}}_0&=\Delta\hat{a}^\dagger\hat{a}+\omega_m\hat{b}^\dagger\hat{b}-g_0\hat{a}^\dagger\hat{a}\left(\hat{b}^\dagger+\hat{b}\right)+\chi\left(\hat{a}^\dagger\hat{a}\right)^2,\\
		\hat{\mathcal{H}}_d&=\mathrm{i}\xi\left(\hat{a}^\dagger-\hat{a}\right),
	\end{split}
\end{equation}
where $\Delta=\omega_c-\omega_d$ is the detuning between cavity mode and the driving field, and $\hat{a}$ ($\hat{b}$) and $\hat{a}^{\dagger}$ ($\hat{b}^{\dagger}$) are the annihilation and creation operator of the quantized field (mirror) respectively. We have defined an {\it effective} frequency $\omega_c=\omega-\chi$, where $\omega$ is the quantized field frequency and $\omega_m$ is the mechanical oscillator frequency. We can express the position operator of the mechanical oscillator in terms of the creation and annihilation operators such that: $\hat{x}=x_{\mathrm{zpf}}(\hat{b}^\dagger+\hat{b})$, where $x_\mathrm{zpf}$ is the zero-point field fluctuation, and $g_0$ is the single-photon coupling strength. To improve the optomechanical interaction and increase the number of photons in the cavity, a laser drive term is added, where $\xi$ is the amplitude and $\omega_d$ is the driving frequency.

Let us next define the Glauber displacement operator \cite{Glauber_1963} on both subspaces (cavity mode and mechanical mode),
\begin{equation}
	\begin{split}
		\hat{\mathcal{D}}_{a}(\eta f_{\hat{b}^\dagger\hat{b}})\ket{k}_c\otimes\ket{j}_m&=\ket{\eta f_{j},k}_c\otimes\ket{j}_m\\
		\hat{\mathcal{D}}_{b}(\eta f_{\hat{a}^\dagger\hat{a}})\ket{k}_c\otimes\ket{j}_m&=\ket{k}_c\otimes\ket{\eta f_{k},j}_m,\\
	\end{split}
\end{equation}
where $\hat{\mathcal{D}}_{\hat{a}}(\eta f_{\hat{b}^\dagger\hat{b}})=e^{\eta f_{\hat{b}^\dagger\hat{b}}(\hat{a}^\dagger-\hat{a})}$, and similarly for $\hat{\mathcal{D}}_{\hat{b}}(\eta f_{\hat{a}^\dagger\hat{a}})$. The set of kets $\{\ket{\alpha f_{j},k},\ket{\alpha f_{k},j}\}$ is recognized as displaced Fock states \cite{Oliveira_1990}, while $\{f_{k},f_{j}\}$ is a set of functions of the photonic or phononic number operator. This operator, also known as the {\it polaron} transformation~\cite{Mahan_2000}, generates phononic displaced number states that in turn depend on the number of photons, and {\it vice versa}. We apply the unitary transformation defined by $\hat{\mathcal{D}}_{\hat{b}}(\eta\hat{a}^\dagger\hat{a})$ to the Hamiltonian $\hat{\mathcal{H}}$, namely $\hat{\mathcal{H}}_{\mathcal{D}}=\hat{\mathcal{D}}_{\hat{b}}^\dagger(\eta\hat{a}^\dagger\hat{a})\hat{\mathcal{H}}\hat{\mathcal{D}}_{\hat{b}}(\eta\hat{a}^\dagger\hat{a})$, and use the following identities
\begin{equation}
    e^{-\eta \hat{a}^\dagger\hat{a}(\hat{b}^\dagger-\hat{b})}\hat{b}e^{\eta \hat{a}^\dagger\hat{a}(\hat{b}^\dagger-\hat{b})}=\hat{b}+\eta \hat{a}^\dagger\hat{a},
\end{equation}
\begin{equation}
    e^{-\eta \hat{a}^\dagger\hat{a}(\hat{b}^\dagger-\hat{b})}\hat{b}^{\dagger}e^{\eta \hat{a}^\dagger\hat{a}(\hat{b}^\dagger-\hat{b})}=\hat{b}^{\dagger}+\eta \hat{a}^\dagger\hat{a},
\end{equation}
\begin{eqnarray}\nonumber
    e^{-\eta \hat{a}^\dagger\hat{a}(\hat{b}^\dagger-\hat{b})}\hat{a}e^{\eta \hat{a}^\dagger\hat{a}(\hat{b}^\dagger-\hat{b})}&=&\hat{a}e^{\eta (\hat{b}^\dagger-\hat{b})}\\&=&\hat{a}\hat{D}_{\hat{b}}(\eta),
\end{eqnarray}
and
\begin{eqnarray}\nonumber
    e^{-\eta \hat{a}^\dagger\hat{a}(\hat{b}^\dagger-\hat{b})}\hat{a}^{\dagger}e^{\eta \hat{a}^\dagger\hat{a}(\hat{b}^\dagger-\hat{b})}&=&\hat{a}^{\dagger}e^{-\eta(\hat{b}^\dagger-\hat{b})}\\&=&\hat{a}^{\dagger}\hat{D}_{\hat{b}}^{\dagger}{(\eta)},
\end{eqnarray}
so we write the displaced Hamiltonian
\begin{eqnarray}\label{H_D0}\nonumber
		\hat{\mathcal{H}}_\mathcal{D}&=&\Delta\hat{a}^\dagger\hat{a}+\omega_m[\hat{b}^\dagger\hat{b}+\eta^2(\hat{a}^\dagger\hat{a})^2+\eta\hat{a}^\dagger\hat{a}(\hat{b}^\dagger+\hat{b})]\\ \nonumber &-&g_0\hat{a}^\dagger\hat{a}\left(\hat{b}^\dagger+\hat{b}+2\eta\hat{a}^\dagger\hat{a}\right)+\chi\left(\hat{a}^\dagger\hat{a}\right)^2\\&+&\mathrm{i}\xi\left[\hat{a}^\dagger \hat{\mathcal{D}}_{\hat{b}}^\dagger(\eta)-\hat{a} \hat{\mathcal{D}}_{\hat{b}}(\eta) \right].
\end{eqnarray}

By setting $\left\{\eta,\chi\right\}=\left\{\frac{g_0}{\omega_m},\frac{g_0^2}{\omega_m}\right\}$ we obtain the ion-laser like interaction
\begin{equation}
	\begin{split}\label{H_D}
		\hat{\mathcal{H}}_\mathcal{D}=\Delta\hat{a}^\dagger\hat{a}+\omega_m\hat{b}^\dagger\hat{b}+\mathrm{i}\xi\left[\hat{a}^\dagger \hat{\mathcal{D}}_{\hat{b}}^\dagger(\eta)-\hat{a} \hat{\mathcal{D}}_{\hat{b}}(\eta) \right].
	\end{split}
\end{equation}
Following this procedure an equivalence has been found between the optomechanical system and Kerr medium \cite{Aldana_2013}, since: $\hat{\mathcal{D}}_{\hat{b}}^\dagger(\eta\hat{a}^\dagger\hat{a})\hat{\mathcal{H}}_0\hat{\mathcal{D}}_{b}(\eta\hat{a}^\dagger\hat{a})=\Delta\hat{a}^\dagger\hat{a}+\omega_m\hat{b}^\dagger\hat{b}$.

With respect to this transformed Hamiltonian \eqref{H_D} the cavity mode plays the role of pseudo-spin operators, {\it i.e.,} this Hamiltonian mimics the ion-laser interaction  \cite{Moya_2000,Ramos_2017,Casanova_2018}, which we write for completeness
\begin{equation}
	\begin{split}\label{H_ion}
		\hat{\mathcal{H}}_{ion}=\nu\hat{a}^\dagger\hat{a}+\frac{\delta}{2}\hat{\sigma}_z+g\left[\hat{\sigma}_- \hat{{D}}^\dagger(i\eta)+\hat{\sigma}_+ \hat{{D}}(i\eta) \right],
	\end{split}
\end{equation}
where the annihilation and creation operators refere to the trap frequency, the Pauli matrices, $\sigma$'s, are the external degrees of freedom operators, $\nu$ is the trapping frequency, $\delta$ the detuning that allows red and blue sideband transitions while $g$ is the laser intensity also called Rabi frequency.

This result is one of our main contributions given that tuning the Kerr medium with the effective Kerr interaction, $(\hat{a}^\dagger\hat{a})^2$ seen in equation (\ref{H_D0})  eliminates the former by simply setting $\chi=\frac{g_0^2}{\omega_m}$, leaving, for some parameters, a soluble Hamiltonian. From here, we show that two quantum systems can be found: a) blue and red sideband interactions and b) two coupled harmonic oscillators.

\section{Generating non-linear mirror-field interactions}
Let us now consider blue and red sideband processes: $\Delta=n\omega_m$ (with $n\in \mathbb{Z}$). For this case the interaction Hamiltonian can be expanded as
\begin{equation}
	\hat{\mathcal{H}}_I=\mathrm{i}\xi e^{-\frac{\eta^2}{2}}\bigg[\hat{a}^\dagger\sum_{j,k=0}^{\infty}\frac{(-1)^j}{j!k!}\eta^{j+k}\hat{b}^{\dagger j}\hat{b}^{k}e^{\mathrm{i}\omega_mt(j-k+n)}-\mathrm{h.c.}\bigg],
\end{equation}
where  we have used the Taylor series for each displacement operator $\hat{\mathcal{D}}_{\hat{b}}(\eta)=e^{\eta(\hat{b}^\dagger-\hat{b})}$. Considering $\xi\ll \omega_m$ we can invoke the rotating wave approximation (RWA), thus conserving the time independent terms {\it i.e.,} $k=j+n$,
\begin{equation}
	\hat{\mathcal{H}}_I=\mathrm{i}\xi e^{-\frac{\eta^2}{2}} \sum_{j=0}^{\infty}\frac{(-1)^j\eta^{2j+n}}{j!(j+n)!}[ \hat{a}^\dagger\hat{b}^{\dagger j}\hat{b}^{j}\hat{b}^n-\hat{a}\hat{b}^{\dagger n}\hat{b}^{\dagger j}\hat{b}^j].
\end{equation}
Using the identity $\hat{b}^{\dagger j}\hat{b}^j=\frac{\hat{N}!}{(\hat{N}-j)!}$ (with $\hat{N}=\hat{b}^\dagger\hat{b}$), we rewrite the above equation as:
\begin{equation}\label{H_IT}
    \begin{split}
	\hat{\mathcal{H}}_I=\mathrm{i}\xi\eta^n e^{-\frac{\eta^2}{2}}&\bigg[\frac{\hat{N}!}{(\hat{N}+n)!}L_{\hat{N}}^{(n)}(\eta^2)\hat{b}^n\hat{a}^{\dagger}\\& -\hat{a}\hat{b}^{\dagger n}\frac{\hat{N}!}{(\hat{N}+n)!}L_{\hat{N}}^{(n)}(\eta^2)\bigg],
    \end{split}
\end{equation}
where $L_{l}^{(n)}(x)$ is the $l$-th Laguerre generalized polynomial with parameter $n$ \cite{NIST}. The above equation shows that a loss of one photon of the field may give rise to $n$ excitations of the mirror and {\it vice versa}. We may dub this the red-sideband excitation process in analogy to processes that take place in the ion-laser interaction. The cavity mode plays the role of pseudo-spin operators. On the other hand, we could consider $n$ to be  a negative integer in which case the loss of a photon produces the loss of $n$ excitations in the mirror states and {\it vice versa}, namely a blue-sideband process \cite{Matos_1996,Wallentowitz_1996,Moya_2000}.

In order to find analytic solutions we consider the on-resonance case $n=0$, thus reducing the interaction Hamiltonian to the simpler form
\begin{equation}\label{H_I}
	\hat{\mathcal{H}}_I=\mathrm{i}\xi e^{-\frac{\eta^2}{2}}L_{\hat{N}}(\eta^2)(\hat{a}^{\dagger} -\hat{a}),
\end{equation}
where $L_j^{(0)}(x)=L_{j}(x)$ is the Laguerre polynomial of order $j$, and whose evolution operator can be written as:
\begin{equation}\label{Df}
\hat{\mathcal{D}}_{\hat{a}}\left[f_{\hat{N}}(t)\right]=\exp\left[f_{\hat{N}}(t)\left(\hat{a}^\dagger-\hat{a}\right)\right],
\end{equation}
with $f_{\hat{N}}(t)=\xi t e^{-\frac{\eta^2}{2}}L_{\hat{N}}(\eta^2)$. As we show below, this function signals non-linearities for both the cavity and the mechanical oscillator.  The Hamiltonians \eqref{H_IT} and \eqref{H_I}, correspond to the ion-laser interaction form of the system defined by \eqref{H_D}.

\subsection{Evolution of the Husimi-Q function}
Let us now compute the state vector  evolution for the initial case of a coherent state both in the cavity and the mechanical oscillator of the optomechanical system: $\ket{\Psi(0)}_{\mathcal{R}}=\ket{\alpha}_c\otimes\ket{\Gamma}_m$. In the rotating frame $e^{-\mathrm{i}\omega_d t\hat{a}^\dagger\hat{a}}$ and $e^{-\mathrm{i}\omega_mt\hat{b}^\dagger\hat{b}}$, we get
\begin{equation}
    \begin{split}
        &\ket{\Psi(t)}_{\mathcal{R}}=\hat{\mathcal{D}}_{\hat{b}}(\eta\hat{a}^\dagger\hat{a})\hat{\mathcal{D}}_{\hat{a}}\left[f_{\hat{N}}(t)\right]\hat{\mathcal{D}}^\dagger_{\hat{b}}(\eta\hat{a}^\dagger\hat{a})\ket{\alpha,\Gamma},\\
        &=e^{-|\alpha|^2}\sum_{k}\frac{\alpha^k}{\sqrt{k!}}\hat{\mathcal{D}}_{\hat{b}}(\eta\hat{a}^\dagger\hat{a})\hat{\mathcal{D}}_{\hat{a}}\left[f_{\hat{N}}(t)\right]\ket{k, \Gamma-\eta k},\\
        &=e^{-|\alpha|^2}\sum_{k,j}\frac{\tilde{\alpha}^k}{\sqrt{k!}}\frac{\tilde{\Gamma}_{k}^j}{\sqrt{j!}}e^{-\frac{|\tilde{\Gamma}_k|^2}{2}}\hat{\mathcal{D}}_{\hat{b}}(\eta\hat{a}^\dagger\hat{a})\hat{\mathcal{D}}_{\hat{a}}\left(f_j\right)\ket{k,j},\\
    \end{split}
\end{equation}
with $\tilde{\alpha}=\alpha e^{\frac{\eta}{2}(\Gamma-\Gamma^*)}$, and $\tilde{\Gamma}_k=\Gamma-\eta k$. In general, we can compute the partially traced Husimi-Q function for either of the two subsystems: $Q^{(i)}(\beta)=\frac{1}{\pi}\braket{\beta|\hat{\rho}_{i}(t)|\beta}$, with $\hat{\rho}_m(t)=\mathrm{Tr}_{c}\left[\hat{\rho}_{\mathcal{R}}(t)\right]$ or $ \hat{\rho}_c(t)=\mathrm{Tr}_{m}\left[\hat{\rho}_{\mathcal{R}}(t)\right]$, respectively. For the mechanical oscillator part, we find:
\begin{equation}\label{Qm}
\begin{split}
        Q^{(m)}&(\beta;t)=\\
        &\frac{e^{-|\alpha|^2}}{\pi}\sum_{n=0}^\infty\bigg|\sum_{k,j=0}^\infty\frac{\tilde{\alpha}^k}{\sqrt{k!}}\braket{n|\hat{\mathcal{D}}_{\hat{a}}(f_j)|k}\braket{j|\tilde{\Gamma}_k}\braket{\tilde{\beta}_n|j} \bigg|^2,
\end{split}
\end{equation}
with $\tilde{\beta}_n=\beta-\eta n$. For the cavity we obtain:
\begin{equation}
    \begin{split}
        &Q^{(c)}(\beta;t)=\frac{e^{-|\alpha|^2}}{\pi}\times\\
        &
	\sum_{l=0}^\infty\bigg|\sum_{k,j,n=0}^\infty\frac{\tilde{\alpha}^k}{\sqrt{k!}}\braket{l|\hat{\mathcal{D}}_{\hat{b}}(\eta n)|j}\braket{n|\hat{\mathcal{D}}_{\hat{a}}(f_j)|k}\braket{j|\tilde{\Gamma}_k}\braket{\beta|n}\bigg|^2.
    \end{split}
\end{equation}
In the following, we concentrate on the reduced density matrix $\hat{\rho}_{m}(t)$ for the mechanical oscillator.  In order to optimize its computational calculation, we take into account that the matrix elements of the displaced  Fock states can be written as:
\begin{equation}
    \braket{n|\hat{\mathcal{D}}(f)|k}=e^{-\frac{|f|^2}{2}}f^{n-k}\frac{(-1)^k}{\sqrt{n!k!}}U\left[-k,n-k+1,|f|^2\right],
\end{equation}
where $U[a,b,z]$ is the Tricomi function, also known as the confluent hypergeometric function of the second kind. This includes the cases for the associated Laguerre polynomials when $n\leq k$, and $n\geq k$.

\begin{figure}
    \begin{center}
        \includegraphics[width=\linewidth]{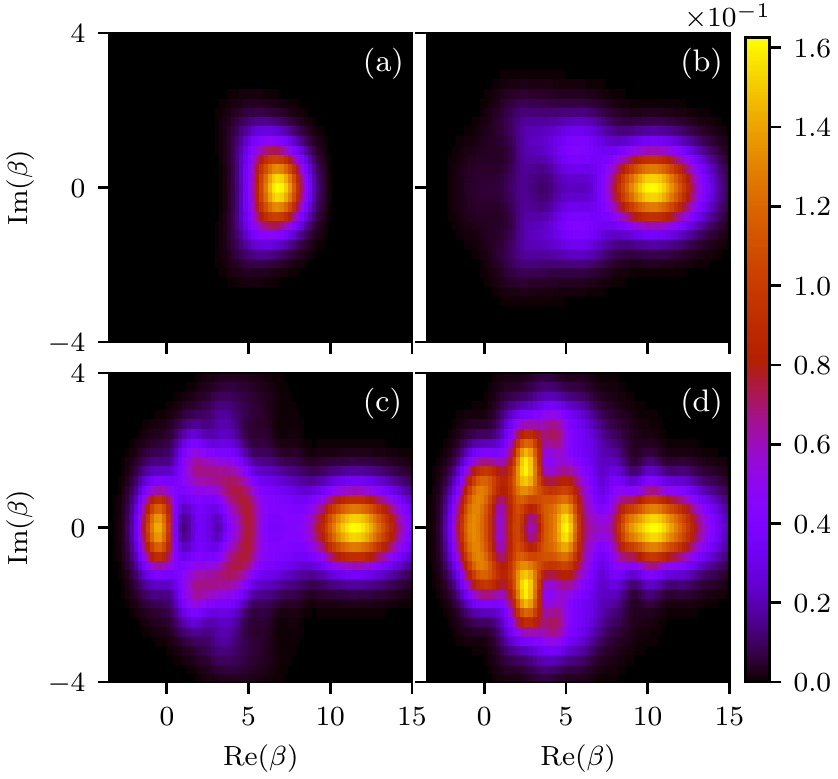}
        \caption{Husimi-Q function for $\ket{\Psi(0)}_{\mathcal{R}}=\ket{\alpha,\Gamma}$, with $\left\{\alpha,\Gamma\right\}=\left\{2,2\right\}$ at scaled time $\xi t=\pi$. In each subfigure from (a) to (c) the ratio is $\eta=\frac{g_0}{\omega_m}=\left\{.25,.5,.75,1\right\}$, respectively, and we also normalize each mesh to the maximum possible value of mesh (a) .  To ensure the convergence of \eqref{Qm}, we have imposed the condition $\int Q^{(m)}(\beta;t)d\beta^2\approx1$ numerically.}
        \label{Fig_2}
    \end{center}
\end{figure}

In Fig.~\ref{Fig_2} we present the dynamic plot of the Husimi-Q function in phase space for \eqref{Qm}. We can easily appreciate that, as we increase the ratio $\eta=\frac{g_0}{\omega_m}$. It is well-known that the interaction between a mirror and a cavity field \cite{Bose_1997} produces several probability spots in phase space when a coherent state is considered as initial state. In our case, where we are considering an extra tuning of a Kerr medium with the system, the nonlinearity coming from the Laguerre polynomials produces a behaviour in the Husimi-$Q$ function as the one shown in Fig.~\ref{Fig_2} where also several spots of probability are found. It is also important to note that, according to the reported experimental parameters, the state of the art of quantum cavity optomechanics, achieving these types of interactions is a challenge that is likely to be achieved in the near future. This behavior is a signature of the intrinsic nonlinearity of the {\it polaron} transformation generating field-displaced Fock states which, in turn, contain a certain number of phonons defined by the Laguerre polynomials $L_{\hat{N}}(\eta^2)$.

\section{Mirror-field interaction as coupled harmonic oscillators}
Let us next study  the case in which  $\Delta\neq n\omega_m$. In this event the RWA cannot be applied. Instead, we consider the weak coupling regime $\eta\ll1$ and we show that the corresponding Hamiltonian \eqref{H_D} transforms into a system of two coupled oscillators:
\begin{equation}
    \begin{split}
	\hat{\mathcal{H}}_\mathcal{D}\approx\Delta\hat{a}^\dagger\hat{a}+\omega_m\hat{b}^\dagger\hat{b}+\mathrm{i}\xi\left[\hat{a}^\dagger-\hat{a}+\eta\left(\hat{a}^\dagger+\hat{a}\right)\left(\hat{b}-\hat{b}^\dagger\right)\right],
    \end{split}
\end{equation}
where we have expanded $\hat{\mathcal{D}}_{\hat{b}}(\pm\eta)$ in its corresponding Taylor series up to first order: $1\pm\eta(\hat{b}^\dagger-\hat{b})$. By defining $\hat{\mathcal{H}}_{\mathcal{THO}}=e^{\mathrm{i}\frac{\pi}{2}\hat{b}^\dagger\hat{b}}\hat{\mathcal{D}}_{\hat{a}}(\alpha)\hat{\mathcal{H}}_\mathcal{D}\hat{\mathcal{D}}_{\hat{a}}^\dagger(\alpha)e^{-\mathrm{i}\frac{\pi}{2}\hat{b}^\dagger\hat{b}}$, and setting $\alpha=\mathrm{i}\frac{\xi}{\Delta}$, we show that
\begin{equation}\label{THO}
	\hat{\mathcal{H}}_\mathcal{THO}=\Delta\hat{a}^\dagger\hat{a}+\omega_m\hat{b}^\dagger\hat{b}+\xi\eta\left(\hat{a}^\dagger+\hat{a}\right)\left(\hat{b}+\hat{b}^\dagger\right).
\end{equation}
where we have suppressed the zero-point energy $\frac{\sqrt{3}\xi^2}{\Delta}$.  In order to obtain the more familiar position-position form of the interaction, we have used the Fourier transform operator. This Hamiltonian corresponds to the usual interaction of two oscillators and has been solved in the context of classical-quantum analogies \cite{Urzua_2019_b}, time-dependent coupling \cite{Ramos_2020}, and by using tools from simplectic geometry \cite{Bruschi_2021}.

The Hamiltonian in \eqref{THO} determines the second kind of interaction derived from the optomechanical interaction, specifically from the polaron form of the Hamiltonian, \eqref{H_D}.
\section{Conclusions}
We have  shown that, by introducing a Kerr medium in a cavity with a moving mirror and tuning it properly, an ion-laser like interaction is rendered. We can set the parameters in such a way that the resulting effective Hamiltonian can be solved analytically. By the same token, but via a different set of parameters and under the weak coupling approximation, we have established that a pair of coupled harmonic oscillators is recovered \cite{Ramos_2020}. Effects of the environment are known to cause decays of  quantum coherences. It is well known that such effects affect the oscillators (especially when coherent states are not considered) and may be studied through master equations and solved by means of superoperator techniques \cite{Arevalo_1998}.

\section*{Acknowledgments}
J.R. and I.R.P. thank Reyes Garc\'{\i}a for the maintenance of our computers, and acknowledge partial support from Direcci\'on General de Asuntos del Personal Acad\'emico, Universidad Nacional Aut\'onoma de M\'exico (DGAPA UNAM) through project PAPIIT IN 1111119.
 %
\end{document}